\documentclass[12pt]{iopart}

\usepackage{graphicx}
\usepackage{epsf}
\usepackage{longtable}
\usepackage{hyperref}
\usepackage{graphics,epsfig,placeins,subfigure,wrapfig}
\usepackage[usenames]{color}
\usepackage{soul}
\usepackage[T1]{fontenc}
\usepackage[utf8]{inputenc}
\graphicspath{{figures/}}

\def\lesssim{\mathrel{\hbox{\rlap{\hbox{\lower4pt\hbox{$\sim$}}}\hbox{$<$}}}}
\def\gtrsim{\mathrel{\hbox{\rlap{\hbox{\lower4pt\hbox{$\sim$}}}\hbox{$>$}}}}
\def\alt{\mathrel{\hbox{\rlap{\hbox{\lower4pt\hbox{$\sim$}}}\hbox{$<$}}}}
\def\agt{\mathrel{\hbox{\rlap{\hbox{\lower4pt\hbox{$\sim$}}}\hbox{$>$}}}}

\def\gta{\ifmmode {\mathbin{\lower 3pt\hbox   
    {$\,\rlap{\raise 5pt\hbox{$\char'076$}}\mathchar"7218\,$}}}
    \else {${\mathbin{\lower 3pt\hbox
    {$\rlap{\raise 5pt\hbox{$\char'076$}}\mathchar"7218\,$}}}
    $}\fi}
\def\lta{\ifmmode {\,\mathbin{\lower 3pt\hbox   
    {$\,\rlap{\raise 5pt\hbox{$\char'074$}}\mathchar"7218\,$}}}
    \else {${\mathbin{\lower 3pt\hbox
    {$\rlap{\raise 5pt\hbox{$\char'074$}}\mathchar"7218\,$}}}
    $}\fi}

\newcommand{\beq}{\begin{equation}}
\newcommand{\eeq}{\end{equation}}
\newcommand{\bea}{\begin{eqnarray}}
\newcommand{\eea}{\end{eqnarray}}

\definecolor{darkperiwinkle}{RGB}{102, 102, 128}

\newcommand{\NCSA}{National Center for Supercomputing Applications, University of Illinois at Urbana-Champaign, Urbana, Illinois 61801, USA}
\newcommand{\ANCSA}{Department of Astronomy, University of Illinois at Urbana-Champaign, Urbana, Illinois 61801, USA}
\newcommand{\PNCSA}{Department of Physics, University of Illinois at Urbana-Champaign, Urbana, Illinois 61801, USA}

\newcommand{\iCASU}{Illinois Center for Advanced Studies of the Universe, University of Illinois at Urbana-Champaign, Urbana, Illinois, 61801, USA}

\newcommand{\UIB}{Departament de F\'\i{}sica, Universitat de les Illes Balears, IAC3 -- IEEC, Crta. Valldemossa km 7.5, E-07122 Palma, Spain}
\newcommand{\AEI}{Max Planck Institute for Gravitational Physics (Albert Einstein Institute), Am Mühlenberg 1, D-14476 Potsdam, Germany}
\newcommand{\TAPIR}{TAPIR, Walter Burke Institute for Theoretical Physics, California Institute of Technology, Pasadena, California 91125, USA}

\newcommand{\WVU}{Department of Physics and Astronomy, West Virginia University, Morgantown, West Virginia 26506}

\newcommand{\WVUII}{Center for Gravitational Waves and Cosmology, West Virginia University, Chestnut Ridge Research Building, Morgantown, West Virginia 26505}

\definecolor{light-gray}{gray}{0.9}

\def\eadnew#1#2{\address{#2 E-mail: \mailto{#1}}}

\begin{document}

\title[Initial data and eccentricity reduction for binary black hole merger simulations]{Initial Data and Eccentricity Reduction Toolkit for \\ Binary Black Hole Numerical Relativity Waveforms}

\author{
  Sarah Habib$^{1,5*}$
  Antoni Ramos-Buades$^{7}$
  E. A. Huerta$^{1,2,3,4}$
  Sascha Husa$^{6}$
  Roland Haas$^{1}$
  Zachariah Etienne$^{8,9}$}
\address{$^1$ \NCSA}
\address{$^2$ \PNCSA}
\address{$^3$ \ANCSA}
\address{$^4$ \iCASU}
\address{$^5$ \TAPIR}
\address{$^6$ \UIB}
\address{$^7$ \AEI}
\address{$^8$ \WVU}
\address{$^9$ \WVUII}
\eadnew{sarahmh2@illinois.edu}{$^{*}$}

\date{\today}

\begin{abstract}
\noindent The production of numerical relativity waveforms that describe quasi-circular 
binary black hole mergers requires high-quality initial data, and an algorithm to 
iteratively reduce residual eccentricity. To date, these tools remain closed source, or 
in commercial software that prevents 
their use in high performance computing platforms. To address these limitations, and 
to ensure that the broader numerical relativity community has access to these tools, herein 
we provide all the required elements to produce high-quality numerical relativity simulations 
in supercomputer platforms, namely: open source parameter files to numerical 
simulate spinning black hole binaries with 
asymmetric mass-ratios; open source \texttt{Python} tools to produce high-quality initial data for
numerical relativity simulations of spinning black hole binaries on quasi-circular orbits; 
open source \texttt{Python} tools for eccentricity reduction, both as stand-alone software and 
deployed in the \texttt{Einstein Toolkit}'s software infrastructure. This open source toolkit 
fills in a critical void in the literature at a time when numerical relativity has an ever increasing role 
in the study and interpretation of gravitational wave sources. As part of our community building 
efforts, and to streamline and accelerate the use of these resources, 
we provide tutorials that describe, step by step, how to obtain and use these open 
source numerical relativity tools. 
\end{abstract}

\pacs{Valid PACS appear here}
\maketitle

\section{Introduction}
\label{sec:intro}

Numerical relativity~\cite{Pretorius:2005gq,Campanelli:2005dd,Baker:2005vv,Cardoso:2014uka} 
plays a central role in contemporary gravitational wave astrophysics~\cite{Sperhake:2014wpa,Centrella:2010mx,baumgarte_shapiro_2010,Alcubierre_book}. 
The use of numerical relativity waveforms has been essential to develop approximate 
waveform models that are extensively used for gravitational wave detection and parameter estimation~\cite{Hannam:2013oca,Bohe:2016gbl,Khan:2015jqa,Blackman:2017pcm,Husa:2015iqa}. 
The construction of numerical relativity waveforms catalogs~\cite{Mroue:2013xna,Healy:2017psd,Boyle:2019kee,Healy:2019jyf,Huerta:2019oxn,Jani:2016wkt} 
has enabled in-depth analyses of 
the astrophysical properties of gravitational wave sources~\cite{Kumar:2018hml,Abbott:2016apu,Lange:2017wki,Lovelace:2016uwp}. 

As gravitational wave astrophysics continues to probe the gravitational wave 
spectrum~\cite{LIGOScientific:2018mvr,TheLIGOScientific:2017qsa,Abbott:2020uma,LIGOScientific:2020stg,Abbott:2020niy}, 
numerical relativity will be essential to enable and interpret new discoveries, enlighten our 
understanding of the physics of these sources, and provide constraints that 
may further establish general relativity or favor alternatives theories of gravity~\cite{Okounkova:2019zjf,Yunes:2016jcc,Nair:2019iur}. 

Advancing our understanding of gravitational wave sources depends critically on the production 
of high quality numerical relativity 
waveforms that, in the case of binary black hole mergers, span an 8-D parameter space 
that includes mass-ratio, two 3-D vectors that define the individual spin of the binary components, 
and orbital eccentricity, \((q,\,\mathbf{s}_1,\, \mathbf{s}_2,\,e)\), respectively. It is then 
apparent that despite the existence of thousands of numerical relativity waveforms, we need to be 
creative about how to combine them to densely sample these high dimensional signal manifold~\cite{Varma:2018mmi,Rifat:2019ltp}. 
It is also clear that we need to continue producing numerical relativity waveforms 
to describe sources whose parameters are not accurately captured by 
existing approximate waveform models or available numerical relativity waveforms. 

In order to empower the broader numerical relativity community to participate in the construction 
of numerical relativity waveform catalogs, we introduce open source \texttt{Python} libraries that 
have been tested and deployed within the \texttt{Einstein Toolkit}~\cite{ETL:2012CQGra} 
to streamline and accelerate 
these activities. This approach builds upon our previous software development that consisted of open source
\texttt{Python} libraries to post-process numerical relativity data to extract the waveform strain at 
future null infinity~\cite{Johnson:2017oel}. These combined tools provide the required 
end-to-end software infrastructure to utilize the \texttt{Einstein Toolkit} for the 
construction of high-quality numerical relativity waveform catalogs. 

This manuscript is organized as follows. Section~\ref{sec:method} describes our approach to 
construct high-quality initial data, and to post-process the data products of numerical relativity simulations 
to remove residual eccentricity. We put these tools at work in Section~\ref{sec:results}, where we show that 
we can produce nearly circular initial data, and that our method for eccentricity reduction produces 
waveforms with eccentricities of order \({\cal{O}}\sim10^{-4}\) after just one iteration. We summarize this work, 
and outline future research directions in Section~\ref{sec:end}.  We present a tutorial 
that describes how to obtain and use these libraries in~\ref{sec:apex}.


\section{Methods}
\label{sec:method}

In this section we describe the approach followed to produce high-quality initial data for 
binary black hole simulations. Thereafter, we briefly introduce the method used for 
eccentricity reduction. 

\subsection{Initial Data Production}

The first guess for the tangential, $p_t$, and radial, $p_r$, 
components of momenta for the black hole binary system are generated
using techniques presented in~\cite{Ramos-Buades:2018azo} 
and~\cite{Healy:2017zqj}. They extract momenta
components from Hamilton's equations of motion in post-Newtonian (PN)
theory, combined with high-PN-order expressions for the
gravitational-wave flux, $dE_{\rm GW}/dt$, and the tidal energy injected
into the black holes, $dM/dt$.

The Hamiltonian contains orbital~\cite{Buonanno:2005xu}, spin-orbit~\cite{Buonanno:2005xu,Damour:2007nc,Hartung:2011te}, 
spin-spin~\cite{Buonanno:2005xu,Steinhoff:2007mb,Steinhoff:2008ji}, and
spin-spin-spin~\cite{Levi:2014gsa} terms up to and
including 3.5PN order.

The high-PN-order
expressions for $dE_{\rm GW}/dt$ incorporate nonspinning and precessing-spin terms~\cite{Blanchet:2013haa,Ossokine:2015vda}, and are adjusted to account
for the tidal energy injected into the black holes $dM/dt$
~\cite{Brown:2007jx}.

The above expressions were implemented in the open-source,
Python-based \texttt{NRPyPN} software, which is part of \texttt{NRPy+}
(``Python-based code generation for numerical relativity... and
beyond!'')~\cite{Ruchlin2018}. A tutorial for using the software
is given in~\ref{sec:idtutorial} below. In short, the expression
for tangential momentum $p_t(r)$ up to and including 3.5PN order is taken from
~\cite{Ramos-Buades:2018azo} and validated up to 3PN order
against~\cite{Healy:2017zqj}, and up to 3.5PN order
against the original \texttt{Mathematica} notebooks used by
~\cite{Ramos-Buades:2018azo}.

Meanwhile, the expression for radial momentum $p_r$ up to and
including 3.5PN order is derived in \texttt{NRPyPN} as follows. First,
Hamilton's equations of motion imply that 

\beq
\frac{dr}{dt} = \frac{\partial H}{\partial p_r}.
\eeq

Next we Taylor expand $\partial H/\partial p_r$ in powers of
$p_r$, about $p_r=0$, to obtain (to first order in $p_r$):
\beq
p_r \approx \left(\frac{dr}{dt} - \left.\frac{\partial H}{\partial p_r}\right|_{p_r=0} \right) \left( \left.\frac{\partial^2 H}{\partial p_r^2}\right|_{p_r=0} \right)^{-1}\,,
\eeq
where
\beq
\frac{dr}{dt}=\left(\frac{dE_{\rm GW}}{dt}+\frac{dM}{dt}\right)\left[\frac{dH_{\rm circ}}{dr}\right]^{-1}\,,
\eeq
and
\beq
\frac{dH_{\rm circ}(r,p_t(r))}{dr} = \frac{\partial H(p_r=0)}{\partial r} 
+ \frac{\partial H(p_r=0)}{\partial p_t} \frac{\partial p_t}{\partial r}\,,
\eeq
are given explicitly in terms of binary input parameters and $M\Omega$
(as given to 3.5PN order by~\cite{Ramos-Buades:2018azo}).

\subsection{Eccentricity Reduction} 
\label{subsec:Eccred}

The algorithm we describe in this section was introduced in~\cite{Ramos-Buades:2018azo}, 
and was originally developed as a \texttt{Mathematica} notebook. As part of this work, 
we have re-written this eccentricity reduction method using \texttt{Python} libraries, optimized it, and 
tailored it to conduct automated, large-scale, numerical relativity campaigns on 
high performance computing platforms. 

This eccentricity reduction method is applied to remove eccentricity from a 
numerical relativity simulation whose initial data were produced with the method 
described in the previous section. Once the numerical simulation has progressed enough, 
typically between \(500\textrm{M}\) to \(600\textrm{M}\) of evolution, we process the relevant data files, 
as described in~\ref{sec:apex}, to compute correction factors,  $\lambda_t, \lambda_r$, 
of the initial components of the momenta \(p^0_t\) and \(p^0_r\). 

To compute  $\lambda_t, \lambda_r$ we assume 
that oscillations induced as a result of eccentricity in the orbital frequency, $\Omega$, take the form 

\begin{equation}
\mathcal{R}_\Omega = A + B \cos ( \Omega_r t + \Psi)\,,
\label{eq:eq0}
\end{equation}

\noindent where $A$, $B$ and $\Psi$ are coefficients to be determined, and $\Omega_r$ is
the frequency of the radial oscillations. Using the 1PN order quasi-Keplerian 
parametrization~\cite{1985AIHS...43..107D}, we can obtain closed form expressions 
for these correction factors

\begin{equation}
\lambda_t = 1+\left[\frac{B}{4 \Omega _0}-\gamma \frac{B  (3 \eta +1)}{8 r_0 \Omega _0}\right]\cos\Psi,
\label{eq:eq1}
\end{equation}
\begin{equation}
\lambda_r = 1+  \frac{B \eta}{2 r_0^{1/2} \Omega _0 \left| p^0_r\right|}    \left[  1    +\gamma \frac{  1   }{ r_0 } \right]\sin\Psi,
\label{eq:eq2}
\end{equation}

\noindent where $\eta= m_1 m_2/(m_1+m_2)$ is the symmetric mass ratio, 
\((m_1,\,m_2)\) represent the masses of the binary components, $r_0$ 
is the initial orbital separation, and 
$\Omega_0$ is the quasi-circular initial orbital frequency calculated at 3.5PN 
order~\cite{Ramos-Buades:2018azo}. In~\ref{sec:idtutorial} we describe how to use 
simulation data and analytical approximations 
to compute the correction factors \(\lambda_t\) and \(\lambda_r\).



\section{Results}
\label{sec:results}

In this section we combine the tools described above for initial data 
production and eccentricity reduction. We selected three binary black hole 
systems whose properties are described in Table~\ref{tab:one_table}. 
Notice that these systems span three different mass-ratios, \(q\in\{1,\,3\}\), 
and several spinning, non-precessing configurations. 

The results presented in Table~\ref{tab:one_table} show that for all the binary systems 
under consideration, our toolkit produces systems whose initial eccentricities are \(e_0\sim10^{-3}\). 
Furthermore, these eccentricity values are reduced to \(e_0\sim10^{-4}\) after just one 
iteration. In other words, these ready-to-use tools produce high-quality numerical relativity 
waveforms after a minimal number of iterations.

\begin{table}[h]
{\small
 \centering
 \begin{tabular}{|c|l|l|l|l|l|l|l|}
  \hline
  $\textrm{D}\, [\textrm{M}]$ & ($\chi_1^x, \chi_1^y, \chi_1^z$) & ($\chi_2^x, \chi_2^y, \chi_2^z$) & $q$ & \textbf{Iter} \# &$p_r\, [\times10^{-4}]$ & $p_\phi\,[\times10^{-2}]$ & $e_0\,[\times10^{-3}]$ \\
  \hline
 11.0 & (0.0, 0.0, -0.4) & (0.0, 0.0, -0.5) & 1.0 &  \textbf{0} &-8.60 & 9.293 &  2.43 \\
$\textbf{S\_q\_1}$ & & & &   \textbf{1}  &-7.70 & 9.284 & 0.74 \\
  \hline
9.0 & (0.0, 0.0, 0.4) & (0.0, 0.0, -0.5) & 3.0 &  \textbf{0} &-7.50 & 7.652 & 1.70 \\
$\textbf{S\_q\_3}$& & & & \textbf{1} &-6.60 & 7.650 & 0.71\\
  \hline
 \end{tabular}
 \caption{Summary of the astrophysical and orbital parameters of three binary black hole systems used 
 to test our open source toolkit for the production of initial data and eccentricity reduction. Notice that in all cases the 
 initial eccentricity is of order \(e_0\sim{\cal{O}}\left(10^{-3}\right)\) for the zeroth iteration, and it is reduced to \(e_0\sim{\cal{O}}\left(10^{-4}\right)\) after just one iteration.}}
\label{tab:one_table}
\end{table}

\begin{figure}
	\centerline{
	\includegraphics[width=0.55\textwidth]{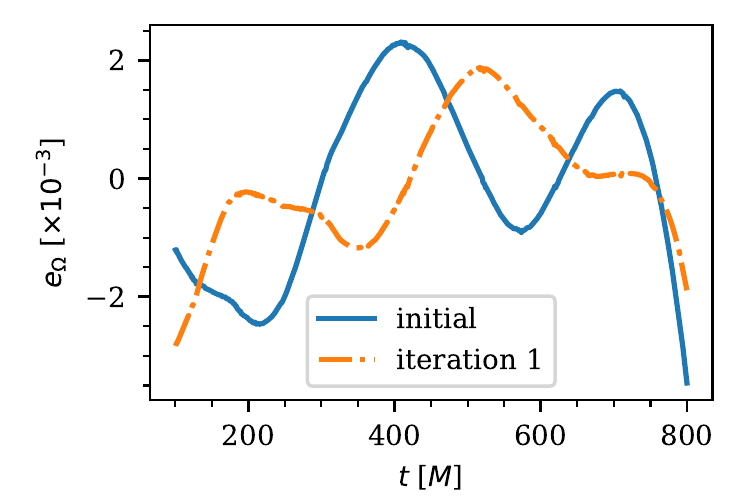}
	\includegraphics[width=0.55\textwidth]{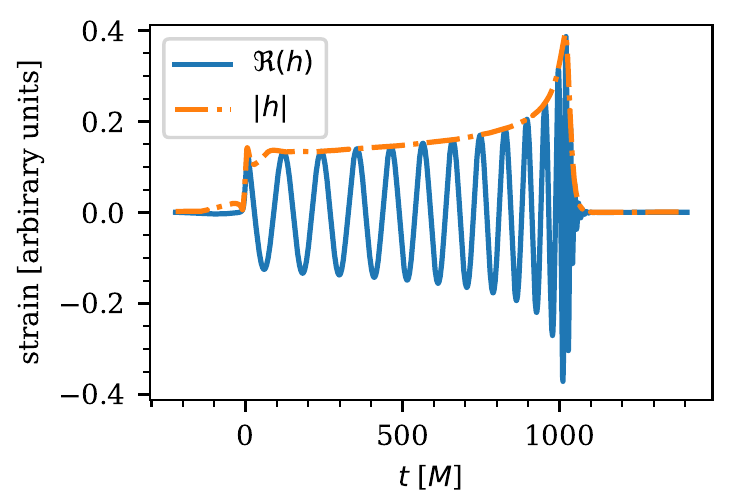}
	}
	\centerline{
	\includegraphics[width=0.55\textwidth]{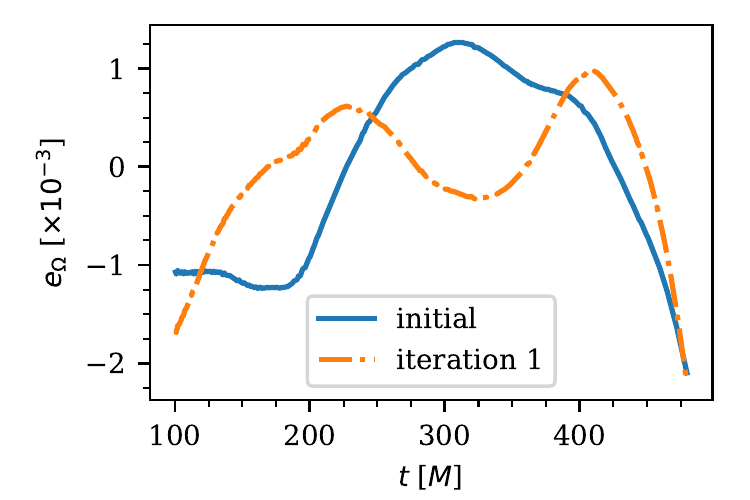}
	\includegraphics[width=0.55\textwidth]{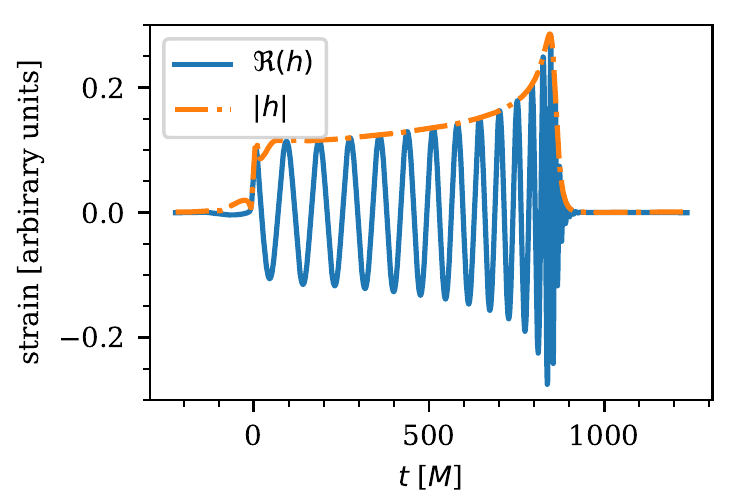}
	}
        \caption{Top panels: results of the eccentricity reduction procedure for simulation $\textbf{S\_q\_1}$. 
        Top-left panel: eccentricity estimator results for the zeroth and
        first iterations. Top-right panel: gravitational wave signal,
        extracted at future null infinity, after eccentricity reduction.
        Bottom panels: as top panels, but for simulation $\textbf{S\_q\_3}$.}
	\label{fig:results1}
\end{figure}

\noindent Figure~\ref{fig:results1} present two types of results. The left panels 
present results for the eccentricity estimator $e_\Omega$  of the orbital
frequency ad defined in Eq.  (3.13) of~\cite{Ramos-Buades:2018azo}. These
results show, as discussed previously, that even the zeroth iteration is already 
nearly circular. The right panels present waveforms of the first iteration  
extracted at future null infinity~\cite{Johnson:2017oel}. 

\noindent These results indicate that the open source tools presented in this article, 
along with the tutorials and configuration files released with this work, will provide 
the required building blocks to engage a broader cross section of the numerical relativity 
community in the production of large scale numerical relativity waveform catalogs. 

\section{Conclusions} 
\label{sec:end}

Numerical relativity simulations~\cite{Pretorius:2005gq,Campanelli:2005dd,Baker:2005vv} 
of binary black hole mergers were produced a decade before 
the first gravitational wave detection of these astrophysical events was realized by 
the advanced LIGO detectors~\cite{LIGOScientific:2018mvr}. Over the last decade, 
numerical relativity software stacks have matured 
to the point of automating and streamlining the production of large-scale numerical relativity 
catalogs~\cite{Mroue:2013xna,Healy:2017psd,Boyle:2019kee,Healy:2019jyf,Huerta:2019oxn,Jani:2016wkt}. 
Nonetheless, the available number of numerical relativity waveforms is not sufficient to densely 
cover the high dimensional signal manifold spanned by these astrophysical events. 

Furthermore, 
essential tools to produce initial data and to automate eccentricity reduction continue to 
be kept as closed source software or licensed software. Neither of these 
solutions is adequate if we aim to enable a larger cross section of the numerical relativity 
community to participate in the production of numerical relativity waveforms to accurately infer 
the astrophysical properties of compact binary sources. This need will become a pressing issue 
as advanced gravitational wave detectors gradually reach design sensitivity, and the number 
of detections reaches the expected number of one event for every fifteen minutes of searched data.

The deployment of these tools as stand-alone software and within the Einstein Toolkit 
is aligned with our community building efforts, 
and marks another milestone in our program for the production of an end-to-end software framework that 
enables users to produce high quality initial data, automate eccentricity reduction, and 
post-process numerical relativity data products to extract numerical relativity waveforms at 
future null infinity~\cite{Johnson:2017oel}. These user-friendly tools will allow new 
users to engage in the development of open source numerical relativity software, using the 
\texttt{Einstein Toolkit} as the driver for such community building activities. 

\section{Acknowledgements} 

\noindent EAH gratefully acknowledges National Science Foundation (NSF) awards 
OAC-1931561 and OAC-1934757. 
RH gratefully acknowledges NSF awards OAC-1550514, OAC-2004879, and ACI-1238993.
This research is part of the Blue Waters 
sustained-petascale computing project, 
which is supported by the 
NSF awards OCI-0725070 and ACI-1238993, and the State of Illinois. Blue Waters is 
a joint effort of the 
University of Illinois at Urbana-Champaign and its National Center for 
Supercomputing Applications. 
We acknowledge support from the NCSA and the Students Pushing INnovation (SPIN) 
undergraduate internship Program at NCSA. 
We thank the \href{http://gravity.ncsa.illinois.edu}{NCSA Gravity Group} for useful feedback. 
NSF-1659702 and XSEDE TG-PHY160053 grants are gratefully acknowledged.

\appendix
\section{Step by Step Tutorial}
\label{sec:apex}

\subsection{Initial Data Production}
\label{sec:idtutorial}

\texttt{NRPyPN} is part of the open-source \texttt{NRPy+}
(``Python-based code generation for numerical relativity... and
beyond!'')~\cite{Ruchlin2018}, and provides the zeroth estimate for
low-eccentricity initial data in this paper. To obtain this estimate
from \texttt{NRPyPN}, first clone the
\texttt{NRPy+} \verb|github| repository: 

{\small
\verb|git clone https://github.com/zachetienne/nrpytutorial.git|
}

Then (assuming that Python 2 or 3 is installed with \verb|pip|),
install \verb|SymPy|~\cite{sympy} and \verb|Jupyter|:

{\small
\begin{verbatim}
pip install -U sympy jupyter
\end{verbatim}
}

Next, from within the \verb|nrpytutorial/NRPyPN/| directory, run

{\small
\begin{verbatim}
jupyter notebook NRPyPN.ipynb
\end{verbatim}
}

A Jupyter notebook will open, in which the binary black hole initial
data parameters for initial separation, spins, and mass ratio can be
specified in the code cell at the bottom of the notebook. When the
code cell is run (\verb|Shift+Enter|), the radial $p_r$ and tangential momenta
$p_t$ to 3.5 post-Newtonian order (largely following~\cite{Ramos-Buades:2018azo}
but fully documented in the linked Jupyter notebooks) will be
output. These momenta can be directly inserted into a Bowen-York
binary black hole initial data solver (the \texttt{TwoPunctures} thorn
was used in this work). For example, for a binary
orbiting in the $xy$-plane with black holes initially located on the
$x$-axis at $x=\pm a$ (with the center of mass at the origin,
$x=y=z=0$), $\pm |p_r|$ will correspond to the $x$-component of
momentum $p_x$ for the puncture at $x = \mp a$, respectively. Also one
may choose $\pm |p_t|$ to correspond to the $y$-component of momentum
$p_y$ for the puncture at $x = \pm a$ or $\mp a$, depending on whether
a clockwise or a counterclockwise orbit is desired.

\subsection{Eccentricity Reduction}

Our eccentricity code is a available as a \texttt{Python3} module on
\verb|GitHub|~\cite{eccred:code}
\small\begin{verbatim}
git clone https://github.com/ncsagravity/eccred
\end{verbatim}\normalsize

Then, assuming Python 3 is installed, run:
\small\begin{verbatim}
python
>>> import EccRed
>>> EccRed.ComputeCorrections("output_glob", MinTime=X, MaxTime=Y)
\end{verbatim}\normalsize
where \textit{output\_glob} is a shell pattern (glob) that matches all
directories containing output files, \verb|MinTime| and \verb|MaxTime|
are time bounds. For best results, \verb|MinTime| should be shortly after any
``junk'' radiation has pass from the vicinity of the black holes and any
initial gauge transition has settled, \verb|MaxTime| should be close to the
time plunge occurs.

Four correction values as well as the estimated eccentricity will be returned
from \verb|EccRed.ComputeCorrections|. In order, they are
$\lambda_r$, $\lambda_t$ computed using two different methods (from PN expansion and from an eccentricity
estimator respectively), and $\delta R$, the
correction factors to radial and tangential momentum components and (additive)
correction to initial orbital separation respectively.
These corrections can then be applied to the respective initial values.

The code expects to two sets of files in the output directories: (i) a file
\textit{TwoPunctures.bbh} as produced by the \texttt{TwoPunctures} thorn that
describes the parameters of the initial black holes, and (ii) a set of
puncture location files \textit{puncturetracker-pt\_loc..asc} as produced by
the \texttt{PunctureTracker} thorn. Columns \textit{pt\_loc\_x[0]},
\textit{pt\_loc\_x[1]} etc., are expected to contain the location of the
original \texttt{plus} and \texttt{minus} punctures. This matches the setup
in~\cite{wardell_barry_2016_155394}.

In the event that \verb|EccRed.ComputeCorrections| throws a runtime error, a likely solution is to adjust \verb|MinTime| or \verb|MaxTime| to better characterize the time domain of inspiral.

\subsection{Automated Eccentricity Reduction}

To simplify automation the process of eccentricity reduction using
\texttt{Simulation Factory}~\cite{simfactory:web} the \texttt{Python} module
can be called as a command line script
\small\begin{verbatim}
./EccRed.py --tmin X --tmax X --input-parfile "input_parameter_file" \
            --output-parfile "output_parameter_file" "output_glob"
\end{verbatim}\normalsize
which automatically applies the correction factors to \texttt{TwoPunctures}'
radial and tangential momentum parameters found in \textit{input\_parameter\_file}
and produces a new parameter file in \textit{output\_parameter\_file}.

We provide a fragment of code in \texttt{RunScript.part} that can be inserted
into \texttt{Simulation Factory}'s run script files to automate the process of
extending a simulation until sufficiently much data has been produced,
estimating eccentricity, computing correction factors, applying them to the
parameter file and submitting a new round of eccentricity reduction.

The fragment contains placeholders \textit{@ECC\_TARGET@} and
\textit{@ECC\_TIME@} for the estimated eccentricity at which to stop the
iteration and the time for which to simulate before applying the correction
algorithm:
\small\begin{verbatim}
sim create --define ECC_TARGET "ECC_TARGET" --define ECC_TIME "ECC_TIME" ...
\end{verbatim}\normalsize
which starts the automated process.

\clearpage

\bibliography{refs}
\bibliographystyle{ieeetr}

\end{document}